\documentclass[12pt]{iopart}

\usepackage{iopams}
\usepackage{graphicx}
\usepackage{color}
\usepackage{siunitx}
\usepackage{booktabs}
\usepackage{multirow}

\newcommand{\ket}[1]{\ensuremath{\left|#1\right\rangle}}

\begin{document}

\title[Interfacing a quantum dot ...]{Interfacing a quantum dot with a spontaneous parametric down-conversion source}

\author{Tobias Huber$^1$, Maximilian Prilm\"uller$^2$, Michael Sehner$^2$, Glenn S Solomon$^1$, Ana Predojevi\'c$^3$ and Gregor Weihs$^2$}

\address{$^1$ Joint Quantum Institute, National Institute of Standards and Technology \& University of Maryland, Gaithersburg, MD 20849, USA}
\address{$^2$ Institut f\"ur Experimentalphysik, Universit\"at Innsbruck, Technikerstr. 25, 6020 Innsbruck, Austria}
\address{$^3$ Institute for Quantum Optics, University Ulm, Albert-Einstein-Allee 11, 89081 Ulm, Germany}
\ead{gregor.weihs@uibk.ac.at}
\vspace{10pt}
\begin{indented}
\item[]\today
\end{indented}

\begin{abstract}
Quantum networks require interfacing stationary and flying qubits. These flying qubits are usually nonclassical states of light. Here we consider two of the leading source technologies for nonclassical light, spontaneous parametric down-conversion and single semiconductor quantum dots. Down-conversion delivers high-grade entangled photon pairs, whereas quantum dots excel at producing single photons. We report on an experiment that joins these two technologies and investigates the conditions under which optimal interference between these dissimilar light sources may be achieved.
\end{abstract}

\vspace{2pc}
%
\noindent\submitto{Quant. Sci. Technol.}

\section{Introduction}
In constructing quantum networks \cite{Kimble08a} light sources for long-distance quantum communications \cite{Briegel98a} play a pivotal role. Semiconductor quantum dots are arguably the best developed single photon source with many advantageous properties: they are mechanically and chemically stable, they don't bleach and thus an individual quantum dot can be used for years. They offer well-understood and characterized energy levels that can be accessed optically and electrically, high quantum efficiency, short radiative lifetimes, and low decoherence at moderate cooling to temperatures of a few Kelvins. With some engineering of the optical environment reasonably high collection efficiencies can be achieved \cite{Somaschi16a,Unsleber16a}. On the other hand, among sources of entangled photon pairs, it appears that spontaneous parametric down-conversion (SPDC), the conversion of a high frequency photon to a pair of lower frequency ones, is still the favorite. While SPDC suffers because it emits a random number of photon pairs and thus usually has to be operated at very low rates, SPDC is usually extremely tuneable in frequency~\cite{Hentschel09a}, and can thus be matched to any quantum memory or stationary qubit. Therefore we find it interesting to try and join these two technologies. Building on earlier results by Polyakov et al.~\cite{Polyakov11a}  we adapt the output of a bright SPDC source to the single photons emitted from an InAs quantum dot embedded in GaAs and demonstrate their Hong-Ou-Mandel (HOM) interference~\cite{Hong87a}. This measurement forms the basis of linear optical Bell-state-analysis~\cite{Michler96a}, quantum teleportation~\cite{Bouwmeester97a} and entanglement swapping~\cite{Pan98a}. These elementary operations feature in the conversion of stationary to flying qubits and in quantum repeater protocols.

Hong-Ou-Mandel interference is the coalescence of two indistinguishable photons into either output port of a beam-splitter. For complete indistinguishability in all photonic degrees of freedom (position-wavevector, time-energy, and polarization) and a perfectly symmetric beamsplitter, two photons that arrive via the two input ports of a beamsplitter, will always exit the beamsplitter together through a common output port, they coalesce. Thus, the detection rate of photons simultaneously exiting from the two different output ports (the so-called coincidence rate) will vanish. For two perfectly indistinguishable photons there are two customary ways of establishing (partial) distinguishability. Either we vary the polarization of one photon, or we apply a time-delay. For delays much larger than the coherence time the two photons become completely distinguishable and behave independently like classical, distinguishable particles. The coincidence rate will be exactly half of the incoming pair count rate, with the other half of the cases resulting in the two distinguishable photons randomly exiting the same output port. This was the method chosen in Ref.~\cite{Hong87a} resulting in a dip in the coincidence rate around zero time delay, the so-called Hong-Ou-Mandel dip. In their experiment Hong et al. used two photons that originated from symmetric SPDC, i.e. two identical infrared photons, and thus the indistinguishability was as good as the spatial modes could be matched on the beamsplitter.

In interfering photons from an SPDC source with those from a quantum dot the challenge of achieving indistinguishability as described before is much bigger. SPDC sources are naturally broad-band whereas quantum dots are rather narrow-line emitters, although not by far as narrow as atomic spectral lines. For interfacing to atomic vapors, or trapped ions resonator-based narrow-linewidth SPDC sources have been built \cite{Wolfgramm08a,Scholz09a,Chuu12a,Luo15a}, but for quantum dots it is sufficient to simply filter the broadband SPDC radiation. When we spectrally filter the photons from an SPDC source that is pumped by a mode-locked laser, it is quite straightforward to achieve a transform-limited wavepacket, whereas decoherence usually degrade the purity of photons emitted by quantum dots. This imperfect purity results in imperfect indistinguishability, even if the central frequencies and linewidths of the two photons are perfectly matched.

In the above-cited experiment~\cite{Polyakov11a} Polyakov et al. attempted the same type of interfacing and achieved a good result. Unfortunately their quantum dot suffered from strong dephasing, which lead to poor indistinguishability of the emitted single photons. In our work, we excited the biexciton state of a quantum dot (two-photon-) resonantly, thus reducing the decoherence and exploiting the much shorter lifetime of the biexciton compared to the exciton. This allowed us to get comparable results with much simpler filtering, which does not require active stabilization.

The article is organized as follows. We start by introducing the two systems, the quantum dot and the SPDC source. Then we introduce the measures taken to match the two types of photons and present the results, which we compare with earlier work. Finally we give an outlook on what would be possible with an improved SPDC source and eventually with different types of quantum dots.

\section{Quantum dot characteristics} \label{sec:qdot}
The quantum dot (QD) sample we used in this experiment contained InAs QDs in GaAs embedded in a $\frac{4\lambda}{n}$ planar microcavity, where $\lambda$ is the quantum dot emission wavelength and $n$ is the refractive index of GaAs. The microcavity consisted of 15.5 lower and 10 upper DBR layer pairs of AlAs and GaAs, each layer being $\lambda$/4 thick. The sample was kept in a helium flow-cryostat temperature stabilized to $\SI{5.0(3)}{K}$ as measured at the cold finger. The excitation pulses originated from a Ti:sapphire laser operating at a wavelength of \SI{918.5}{nm} with a repetition rate of \SI{82}{MHz}, which corresponds to a repetition period $T_L=\SI{12.2}{ns}$.

The photons emitted from a quantum dot show a different degree of indistinguishability depending on how the quantum dot is excited. In particular, resonant excitation to the biexciton enables a high indistinguishability compared with non resonant pumping schemes~\cite{Huber15a}. This is due to two reasons: the resonant excitation creates the biexciton state jitter free~\cite{Flagg12a} and the quantum dot environment does not suffer from electrostatic fluctuations present in the above-band excitation. The latter are caused by charge traps in the vicinity of the quantum dot, which are present even in the best manmade material. Unintentional dopants lead to deep levels that get randomly filled by the free carriers induced by a high-energy excitation laser and are randomly ionized again. Often this unwanted behavior can be reduced by continuously applying a small amount of above-band light, so as to keep these levels filled and thus reduce the fluctuation \cite{Huber16b}.

\begin{figure}[ht]
  \includegraphics[width=0.8\textwidth]{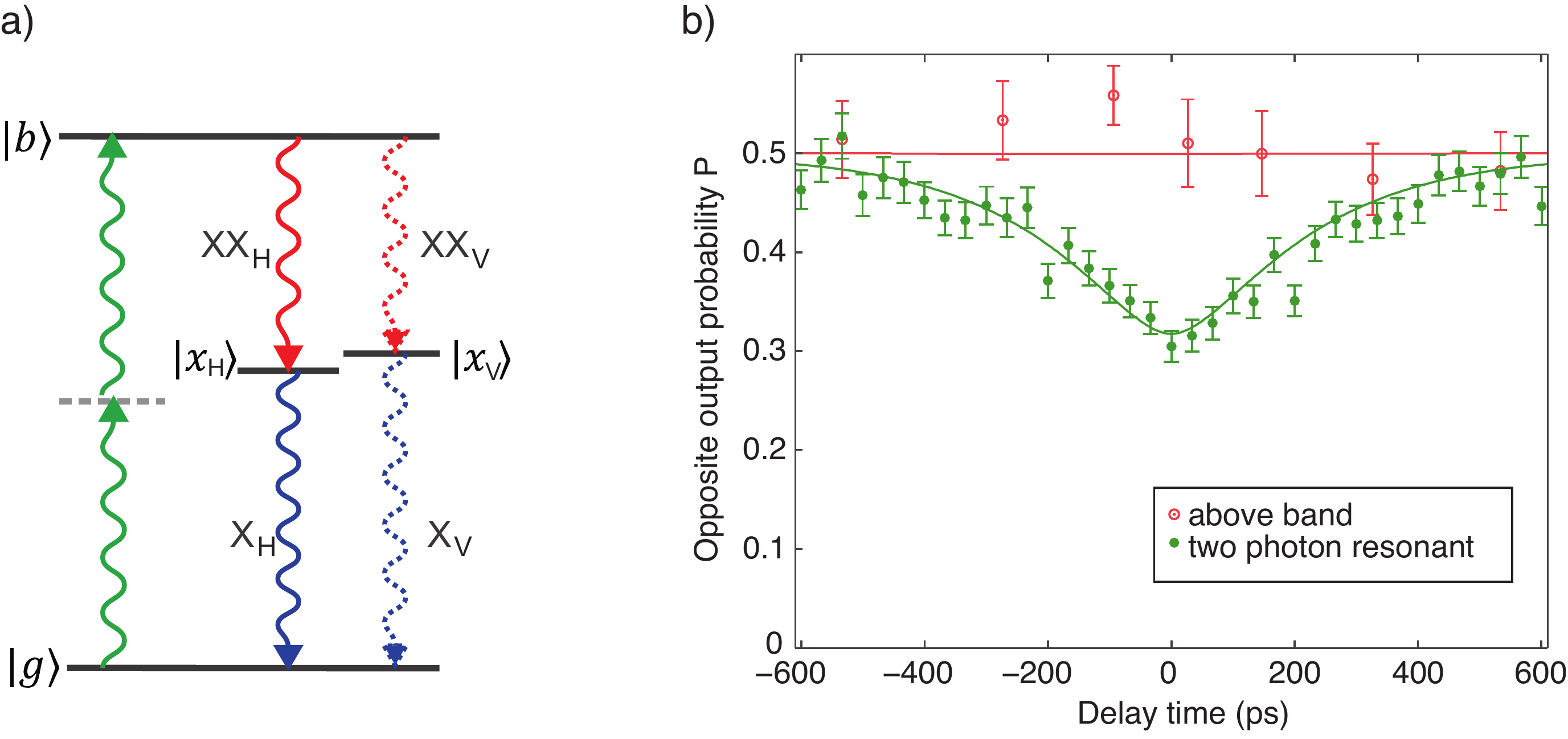}
  \centering
  \caption{\textbf{a)} Energy schematics of the quantum dot. The pump laser (green) is pumping the ground state \ket{g} to biexciton state \ket{b} transition with a two photon process. This results in a cascade of two photons, the biexciton recombination $XX_{H,V}$ followed by the exciton recombination $X_{H,V}$. \textbf{b)} HOM interference from a quantum dot biexciton recombination photon with a replica of itself, created at later times. This plot is based on part of the data previously published in Ref.~\cite{Huber15a}. The time-delay is the time difference of the two photons arriving at the beamsplitter. The red open circles are data points measured with above band (HeNe) excitation, and the green filled circles are data points obtained with two photon resonant excitation.}
  \label{fig:qd_hom}
\end{figure}

The two-photon resonant excitation method we used~\cite{Jayakumar13a} coherently couples the ground state and the biexciton state via a two photon resonance (see Fig.~\ref{fig:qd_hom}a for the energy schematics of a QD). The excitation laser wavelength lies about centered between the biexciton ($\lambda_{XX}=\SI{919.6}{nm}$) and exciton ($\lambda_X=\SI{917.7}{nm}$) luminescence wavelengths. Therefore, any scattered laser light may be rejected from the emitted luminescence photons by spectral filtering. Additionally, we used an orthogonal excitation geometry to further suppress the scattered laser light. This excitation method yields a high photon pair generation efficiency. It also enables the generation of time-bin entanglement~\cite{Jayakumar14a}.

In Ref.~\cite{Huber15a} we reported the indistinguishability of photons consecutively emitted by a quantum dot. The result, plotted in Fig.~\ref{fig:qd_hom}b, shows an indistinguishability contrast of 0.39(2). This indistinguishability was measured using biexciton recombination photons. We showed in Ref.~\cite{Huber15a} that this result can be improved to 0.71(3) by active wavepacket shaping using a modulator. In the task at hand we chose not to apply any active modulation, because the SPDC pulses are very short in time and it is sufficient to employ temporal filtering of the detection events. Both types of filtering can improve the indistinguishability because the QD emission is generally not lifetime limited, i.e., the QD emission is inhomogeneously broadened, which leads to a coherence time $T_2$ which is shorter than twice the lifetime $T_1$.

It has been found \cite{Thoma16a} that the coherence time one extracts from HOM interference measurements between successive photons from a quantum dot is often longer than the one measured with a Michelson interferometer. The likely reason for this is that any slow fluctuations such as those of the charge environment do not degrade the indistinguishability of successive photons too much if the environment is quasistatic between the two emission events that are only separated in time by a few nanoseconds. On the other hand, in a Michelson interferometer with a long and a short arm, any change in the central frequency (often called spectral diffusion) of the emitted photons will degrade the contrast of the averaged interference pattern and thus result in a shorter coherence time. This is consistent with our findings of Ref.~\cite{Huber15a}, where we extracted a significantly longer coherence time from the HOM measurement than from the Michelson measurement. For the case at hand, the HOM interference between quantum dot and SPDC photons we expect the effect to depend on the bandwidth of the SPDC. For extremely narrow filtering the drift of the central frequency should have worse effect than for somewhat wider-band SPDC radiation. Nevertheless, as HOM interference is not intrinsically phase-sensitive the effect of spectral diffusion will be less pronounced, even though the apparent lengthening of the coherence time will still be observable.

\section{SPDC characteristics}
We built the spontaneous parametric down-conversion (SPDC) source in linear geometry. The nonlinear crystal we used was a \SI{15}{mm} long type II periodically poled potassium titanyl phosphate (ppKTP) crystal, where the poling period (\SI{26.4}{\micro\meter}) was chosen for down-conversion of a single \SI{460}{nm} photon into two \SI{920}{nm} photons. By changing the temperature it was possible to tune the down-converted photons to be non-degenerate. We exploited this tunability to match the central wavelength of the signal photons to the quantum dot biexciton photon wavelength of \SI{919.6}{nm}. The SPDC source was pumped with the same pulsed Ti:sapphire laser as the quantum dot to generate photons synchronously. Since the SPDC source required a pump near \SI{460}{nm}, we used second harmonic generation in a beta-barium borate (BBO) crystal to double the laser frequency.

\begin{figure}%
  \centering
  \includegraphics[width=0.5\textwidth]{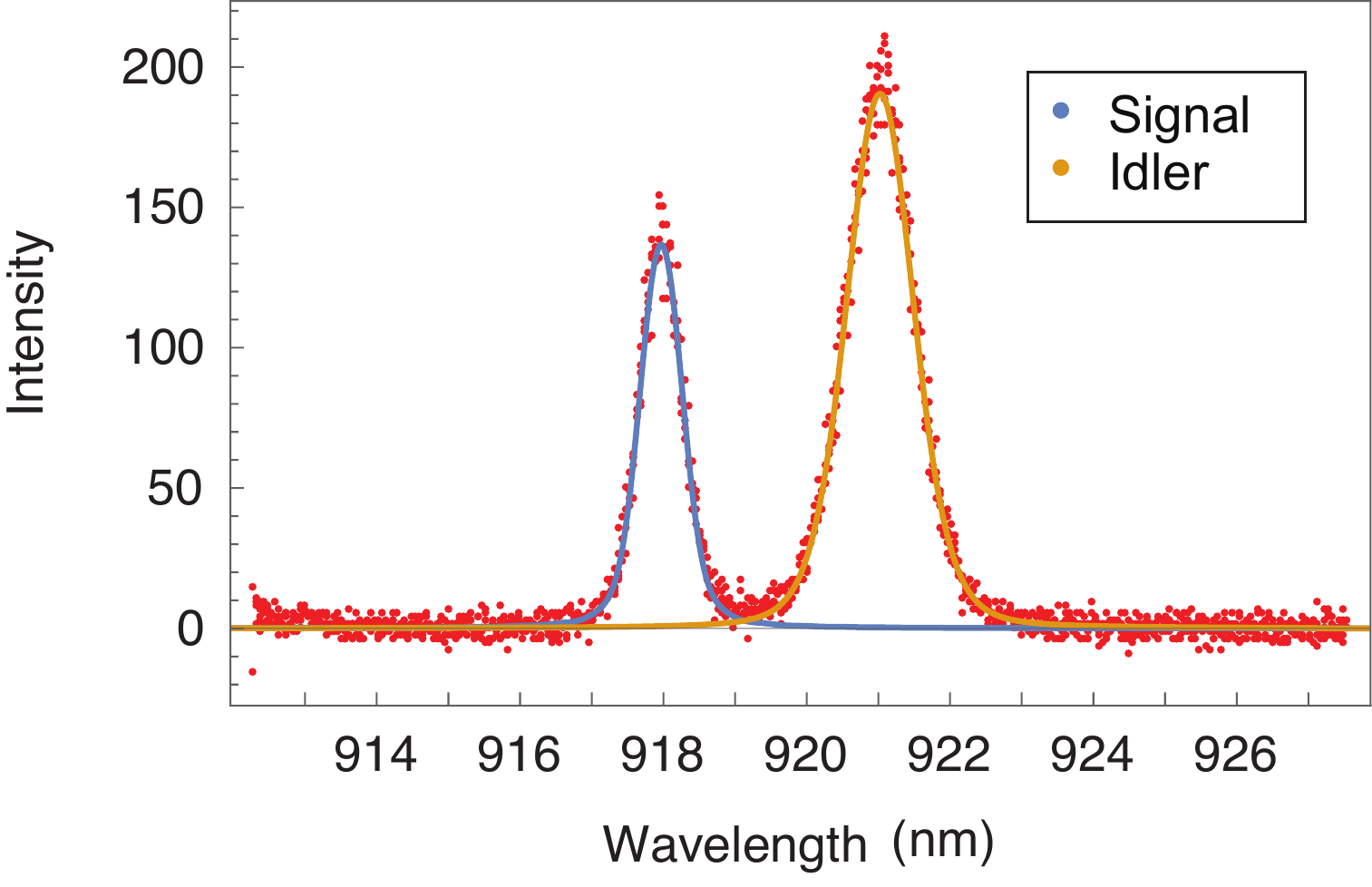}
  \caption{Spectrum of signal and idler of the SPDC source together with the theoretical model for a \SI{3}{ps} pump pulse.}
  \label{fig:spdc_spectrum}
\end{figure}

The joint spectrum of the two SPDC photons in a pair is governed by energy conservation and phase matching. Energy conservation dictates that the frequencies of the signal and idler photons must sum to a frequency that lies within the pump laser spectral line. Phase matching is the approximate conservation of photon momentum and thus depends on the crystal refractive indices for the various involved modes. As a result the joint spectrum is often highly anti-correlated with a broad distribution of frequencies for either photon individually, but a more or less narrow sinc-shaped bandwidth for the sum frequency \cite{Grice01a}. The measured spectra of signal and idler of our SPDC source for the nondegenerate case together with the theory curves are shown in Fig.~\ref{fig:spdc_spectrum}. Through the measurement of one photon, we can, in principle, determine the frequency of the other photon with high accuracy via the strong anti-correlation. In turn this means that each photon's individual state (traced over the partner photon's frequencies) is in a mixed state and the photons have poor indistinguishability, leading to poor interference with independently created, other photons. In our experiment, SPDC is used as a heralded single photon source. While the signal arm of the SPDC source will interfere with the quantum dot photons, the partner SPDC photon (idler) serves as a trigger or herald for the presence of the other one. The heralding efficiency of the unfiltered source was $9.2\%$. For interference with independently created photons narrow spectral filtering has to remove the frequency anticorrelation between the two SPDC photons. After filtering, the photon going to the beamsplitter will individually be in a pure state, fit for interference with the quantum dot photon.

\begin{figure}[ht]
  \centering
  \includegraphics[width=\textwidth]{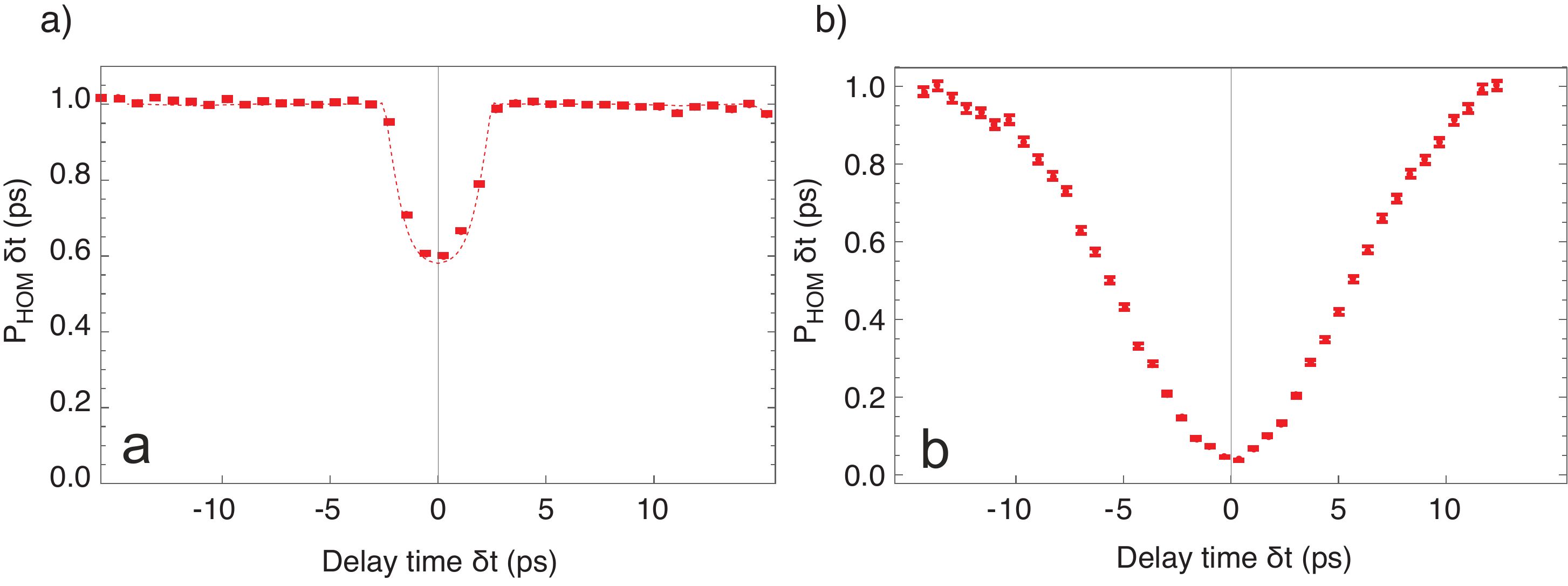}
  \caption{HOM interference measured between the signal and idler photon of the SPDC source.  \textbf{a} shows the unfiltered SPDC for a \SI{6}{ps} pump pulse and no filtering. The statistical error bars are very small and partially cover the symbols. The dashed line is  a fit to the data for better clarity. \textbf{b} shows the same with a \SI{30}{GHz} fiber Bragg grating (FBG) filter in one of the outputs after the 50:50 beamsplitter resulting in a very clean dip of about \SI{96.3(2)}{\%} visibility.}
  \label{fig:spdc_hom}
\end{figure}

To characterize the correlation properties of the down-conversion source itself we performed a HOM interference measurement between the signal and the idler, with the temperature tuned to degeneracy of the source. Due to the narrow phase-matching function, even at degeneracy the signal and idler spectra are not identical but the idler is wider than the signal, which can be seen (for the non-degenerate case) in Fig.~\ref{fig:spdc_spectrum}. Thus the indistinguishability of the unfiltered signal and idler photons was relatively poor resulting in a visibility of only \SI{39.9(4)}{\%} as shown in Fig.~\ref{fig:spdc_hom}a). Restricting the spectral bandwidth through a \SI{30}{GHz} fiber Bragg grating (FBG) filter in one of the output ports of the beamsplitter strongly improved the indistinguishability and thus the HOM interference visibility to \SI{96.3(2)}{\%} (see Fig.~\ref{fig:spdc_hom}b). This happened with an FWHM pump pulse duration of approximately \SI{10}{ps}, which is just a little shorter than the Fourier-transform limit of \SI{14.7}{ps} FWHM corresponding to the \SI{30}{GHz} filter bandwidth. We include these results to demonstrate the effect of the FBG on the SPDC radiation which was used to generate one set of data quantum dot - SPDC HOM interfere reported below.

\section{HOM Interference}
In order to achieve high quality interference between quantum dot emission and down-converted photons, their time/frequency shape is the crucial parameter. While position/wavevector, absolute time/central energy, and polarization can easily be adjusted or tuned, shaping the wavepacket or reducing dephasing is more difficult. The decay of an excited quantum dot state is predominantly radiative and spontaneous, resulting in an exponential time envelope of the emitted photons. The Fourier transform of this exponential decay function is a Lorentzian function in the spectrum. The SPDC photons' individual spectra on the other hand are typically sinc-shaped, as explained above.

\begin{figure}
  \centering
  \includegraphics[width=\textwidth]{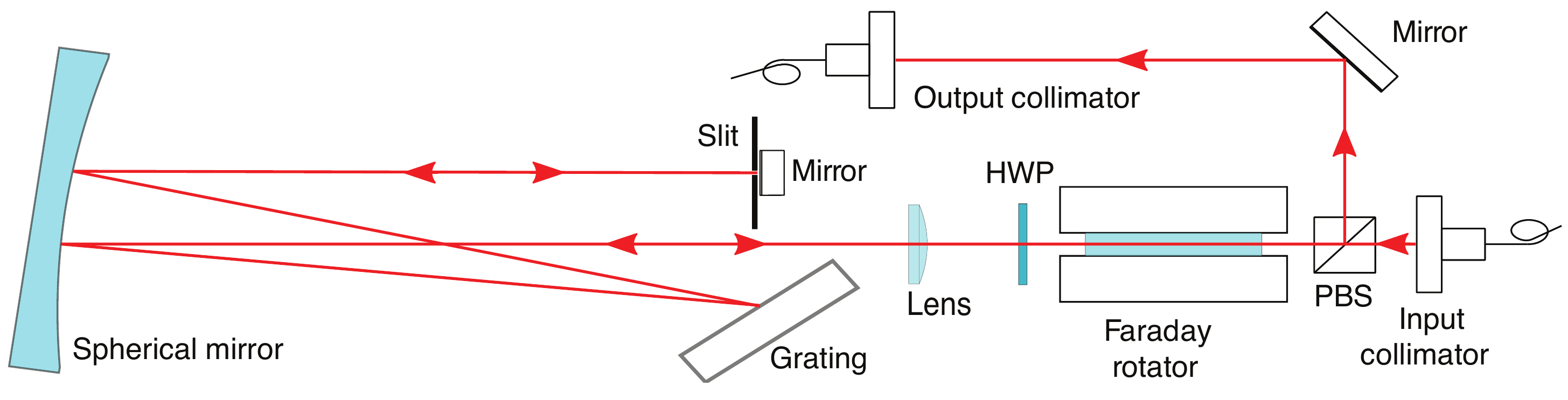}
  \caption{Schematic diagram of the frequency filter. The main elements are a spherical mirror with a focal length of \SI{500}{mm}, a blazed diffraction grating with a grating constand of \SI{1800}{mm^{-1}} and a mechanical slit. The filtered photons were separated from the input by a Faraday isolator.}
  \label{fig:stretcher}
\end{figure}

Besides the different spectral shapes of the photons from the QD and from SPDC, their bandwidth is also significantly different. While the QD emission has a natural width of about \SI{1}{GHz}, the SPDC photons' bandwidth is of the order of \SI{100}{GHz}. To overcome this problem, we filtered the SPDC bandwidth down to \SI{7.7(1)}{GHz} (FWHM), as estimated from a Michelson interference autocorrelation measurement. This was the preactical technical limit of our filtering setup but had the advantage of making active stabilization of the center frequency obsolete. Unfortunately this filtering reduced the (mean) heralding efficiency to \SI{1.5}{\%}. As a filter we used a pulse stretcher configuration of two diffraction gratings in a folded, space-saving 4-$f$ configuration, see Fig.~\ref{fig:stretcher}.

\begin{figure}
  \centering
  \includegraphics[width=0.7\textwidth]{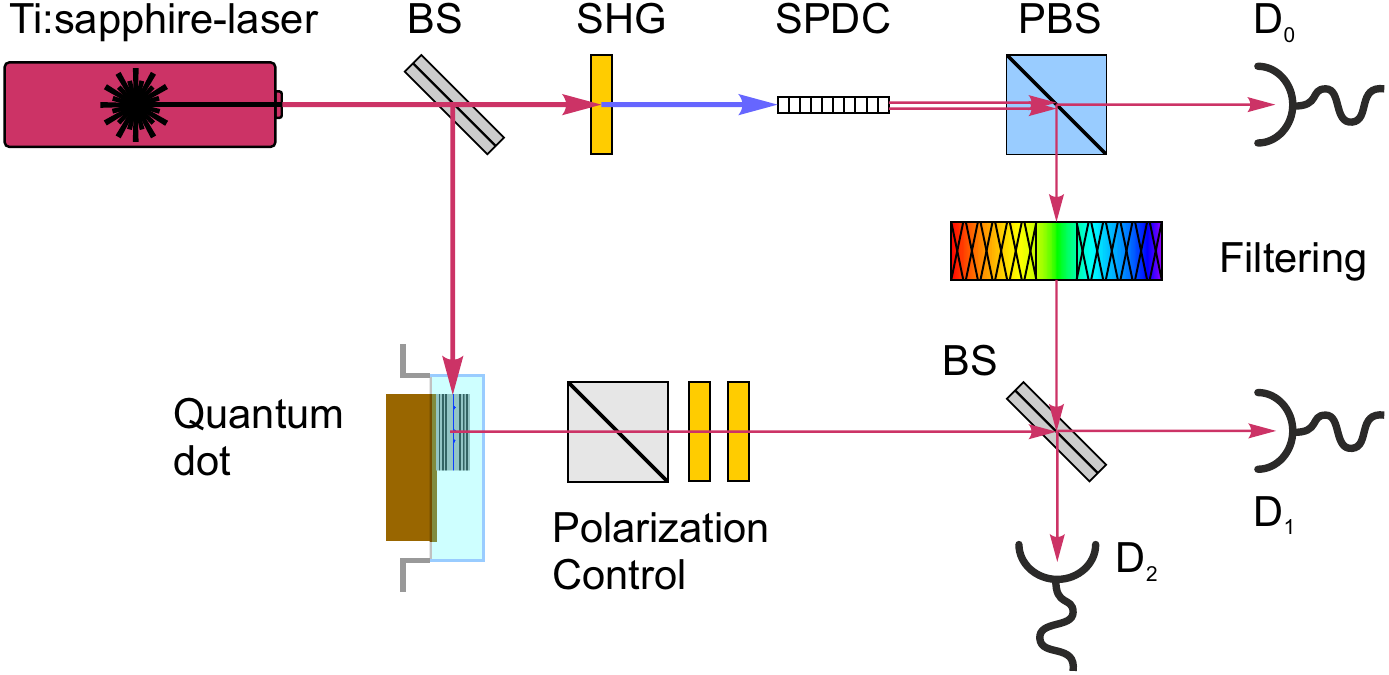}
  \caption{Schematic diagram of the experiment. A variable ratio beamsplitter (BS) divides the laser (Ti:Sapphire) power between the second-harmonic-generation (SHG) and the quantum dot excitation. The frequency-doubled laser light pumps type-II SPDC in a PPKTP crystal to generate photon pairs, which are separated at a polarizing beam-splitter (PBS). One photon (idler) is detected in single photon detector $\mathrm{D_0}$ and serves as a herald, whereas the other is directed towards the filtering setup described in Fig.~\ref{fig:stretcher}. The quantum dot resides in a flow cryostat and is excited through the cleaved wafer edge by a small fraction of the laser power. The photons emitted from the quantum dot are captured by a high-NA lens (not shown), directed through a monochromator (not shown) to pick the biexciton line, and pass a polarization control stage consisting of a Glan-Thompson polarizer and two waveplates. The quantum dot photon and one SPDC photon (signal) meet at the second beamsplitter (BS). Before conducting any measurements we set their relative timing once by changing the path length of the SPDC signal photon via moving the fiber coupler so that the two photons nominally arrive at the beamsplitter simultaneously. The outputs of this beamsplitter are monitored by two single photon detectors $\mathrm{D_1}$ an $\mathrm{D_2}$. All detector signals are routed to a Picoquant Hydraharp time-tagging system (not shown), where the event times are recorded with high time resolution against an electronic trigger signal derived from the laser.}
  \label{fig:hybridhomsetup}
\end{figure}

The entire setup is schematically depicted in Fig.~\ref{fig:hybridhomsetup}. After the filter, the SPDC photons were delayed to match the arrival time of the quantum dot photons on the 50:50 beamsplitter by moving the element that coupled the SPDC signal photons to an optical fiber. This matching was only performed once and not changed thereafter. The laser was a Ti:sapphire laser operating at \SI{918.5}{nm} producing pulses of about \SI{10}{ps} duration and \SI{82}{MHz} repetition rate. A beta barium borate (BBO) crystal frequency-doubled a large part ($\approx\SI{300}{mW}$) of the laser power to serve as a blue pump for the down-conversion process. This blue light at a wavelength of  \SI{459.25}{nm} was cleaned up spatially through a short piece of single mode fiber and the remaining \SI{4.5}{mW} were then focussed onto the ppKTP crystal. The quantum dot was excited with \SI{1.2}{mW} of the \SI{918.5}{nm} light as described in section~\ref{sec:qdot}. Single biexciton photons from the quantum dot and the signal photons from the SPDC source (both at \SI{919.6}{nm} central wavelength) were sent simultaneously towards a 50:50 beamsplitter to observe the Hong-Ou-Mandel interference. Single photon counting avalanche photo-diodes $\mathrm D_0$, $\mathrm D_1$ and $\mathrm D_2$ monitored the idler arm for heralding and the two output ports of the beamsplitter, respectively. A  multi-channel event timer recorded the photon detection times with a resolution of \SI{128}{ps} so that coincidence and time filtering could be implemented in software.

We considered only events where there was a herald photon detected at the heralding detector $\mathrm D_0$ within the same repetition period of the laser, i.e. ``double'' and ``triple'' coincidences. We then calculated the time (micro-time) relative to the laser pulse (macro-time), which served as a synchronization signal via a fast photodiode. This generated a conditioned list of heralded events, which could have detections at only one (``doubles'') or both of the detectors $\mathrm D_1$ and $\mathrm D_2$ (``triples'').

\begin{figure}
  \centering
  \includegraphics[width=\textwidth]{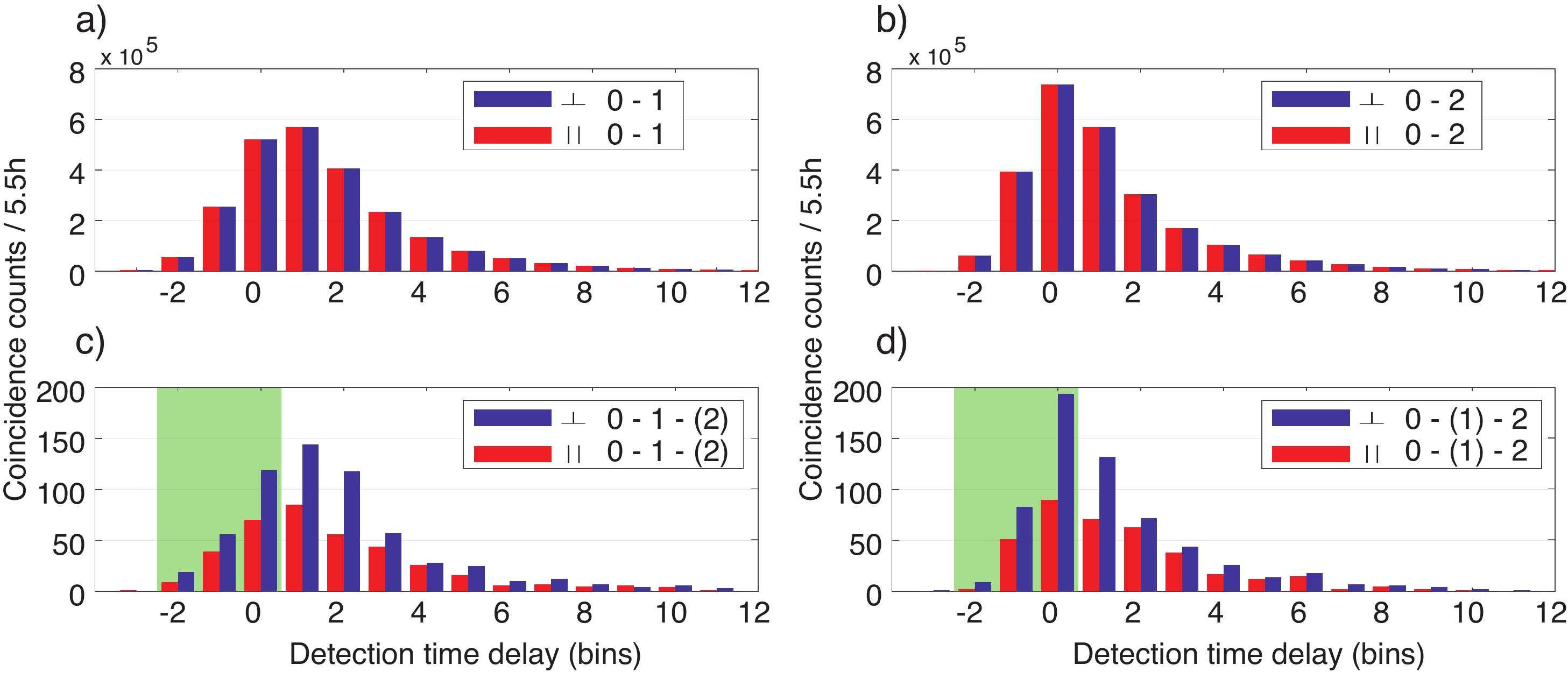}
  \caption{The top two histograms \textbf{a)} and \textbf{b)} summarize photon detection times relative to the synchronizing laser pulse by either detector $\mathrm D_1$ and $\mathrm D_2$, respectively, only for cases when the heralding detector $\mathrm D_0$ detected a photon within the same laser repetition period (``doubles''). Blue bars are for orthogonal input polarizations to the HOM-beamsplitter, red ones for parallel polarizations. The detection times are discretized into \SI{128}{ps} time bins, resulting in a shape that is a convolution of the quantum dot radiative lifetime, the detector response and the discretization. The SPDC pulse is so short that its width should not have any noticeable effects. The histograms show that there is no overall change in the number of detected events or their temporal distribution for the two polarization configurations. The bottom histograms \textbf{c)} and \textbf{d)} show those subsets of data of \textbf{a)} and \textbf{b)}, respectively, in which both $\mathrm D_1$ and $\mathrm D_2$ clicked in the same laser repetition period as the heralding detector $\mathrm D_0$ (``triples''). The temporal shapes of both histograms are consistent with an unbiased sample from the data in \textbf{a)} and \textbf{b)}. The light green shaded areas in \textbf{a)} and \textbf{b)} mark the time bins that we used later for temporal filtering, so as to only count the early and thus coherent photons emitted by the quantum dot.}
  \label{fig:pulsehistograms}
\end{figure}

The distributions of micro-times (Fig.~\ref{fig:pulsehistograms}) for both detectors $\mathrm D_1$ and $\mathrm D_2$ clearly show that the timing and overall event counts are independent of polarization. They also indicate that there are no significant difference between the distributions of ``doubles'' and ``triples'', however the reponses of $\mathrm D_1$ and $\mathrm D_2$ are somewhat different, most likely due to the time discretization in conjunction with slightly different detector signal shape and temporal offset. The fact that the histograms Fig.~\ref{fig:pulsehistograms}c) and d) exhibit reduced event counts for parallel polarizations as compared to the orthogonal case is caused by HOM interference, however, the contrast exhibited by the data here is not the contrast of the actual HOM interference, because in this case the second photon could be in any time bin, i.e. no narrow coincidence window is applied to the time difference between $\mathrm D_1$ and $\mathrm D_2$.

The heralded macro-time list defined its own ordered series of events from which we created a global pseudo-time axis, by assigning one laser repetition period $T_L$ to the (macro-)time difference between one heralding event sync laser pulse and the next, no matter how far apart they were in real time. We did this to be able to assess the correlation or delayed-coincidence of independent events, which serves as a reference value for the HOM interference. To obtain the Hong-Ou-Mandel signals we then took the differences between the pseudo times of $\mathrm D_1$ and $\mathrm D_2$, which of course are real times for actually coincident events, i.e. where both detectors click within the same laser repetition period as the heralding detector (``triples''). We recorded data for \SI{5.5}{h} for each polarization configuration, parallel and orthogonal. This actually happened in several blocks of time, between which various parameters of the laser, sources, and filters were checked and adjusted, if necessary.

\begin{figure}%
\centering
\includegraphics[width=\textwidth]{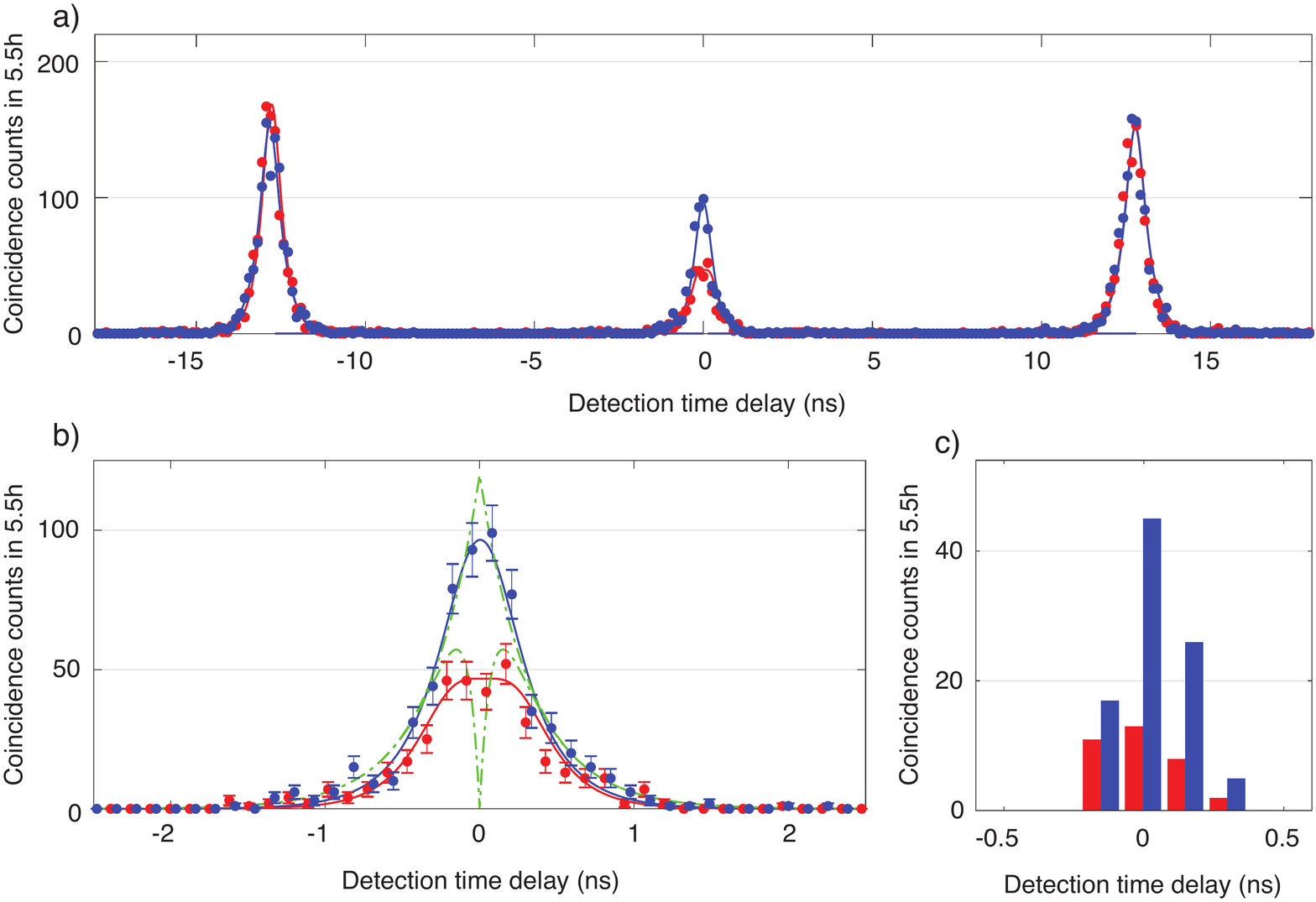}
\caption{All three panels show triple coincidences ($\mathrm D_0$\&$\mathrm D_1$\&$\mathrm D_2$), i.e. heralded HOM interference as a function of the detection time difference $\mathrm D_2-\mathrm D_1$, whereby the timing of the photons into the HOM beamsplitter was kept constant at simultaneity. Coincidence counts are shown as red and blue symbols for parallel and orthogonal polarizations configurations, respectively, with fit curves in matching colors.  The central peak in panel \textbf{a)} shows a significant suppression of coincident events for parallel polarization compared to orthogonal polarization. The side peaks on the other hand do not exhibit any significant change between the two configurations. As explained in the text, the side peaks are displayed on a pseudo-time axis for which subsequent heralding events are offset by one laser pulse repetition period. We omitted the Poissonian error bars here for the sake of visual clarity. Panel \textbf{b)} shows the central peak in higher resolution. The peak shape and fit curves are discussed in the main text as well as the two dashed green lines, which show what we would expect for detectors without any time jitter. Finally, in panel \textbf{c)} we show a histogram of ``triples'' where we selected data from only the first three time bins after the laser pulse (green shaded area in Fig.~\ref{fig:pulsehistograms}c) and d)). }%
\label{fig:hybrid_hom}%
\end{figure}

The resulting correlation signals for orthogonal and parallel polarizations are shown in Fig.~\ref{fig:hybrid_hom}a) on the global pseudo-time axis, where they exhibit peaks at integer multiples of $ T_L$. These side peaks turn out to be identically high for parallel and orthogonal polarizations within the Poissonian experimental uncertainty. Also, as expected the central peak for orthogonally polarized photons is half as high as the side peaks because we are looking at a signal produced by two single photon sources (one heralded) where even for fully distinguishable photons the probability to observe two photons simultaneously is only half of the probability of observing two photons at different times. This distinguishable central peak serves as the reference for the actual HOM measurement.

For further analysis we would like to focus on Fig.~\ref{fig:hybrid_hom}b), which shows the central peaks on an enlarged time axis, again in blue for orthogonal and in red for parallel polarization. The main result is that the data for parallel polarization shows a substantially reduced and flattened peak compared to the orthogonal reference peak. Let us first discuss the features of the orthogonal polarization case. We expect its shape to be governed by the comparably slow quantum dot radiative lifetime $T_1$, with additional broadening by the finite detector time jitter $T_D$, but presumably not significantly  by the SPDC photon. The latter had a frequency bandwidth of \SI{7.7}{GHz} (FWHM) corresponding to a pulse duration of about \SI{95}{ps} (FWHM). We modeled the shape as a convolution of a double sided exponential with $T_1 = \SI{328(7)}{ps}$, as obtained from a separate lifetime measurement and fitted the data by with a Gaussian detector response $T_D=\SI{240}{ps}$ (FWHM) and a free amplitude parameter. The data appear to exhibit some small oscillations in the wings of the peak, which might stem from the SPDC temporal shape which is not purely Gaussian, but also oscillates weakly in the wings as a result from the spectral filtering with a slit. The green dashed lines in Fig.~\ref{fig:hybrid_hom}b) are model curves with the same parameters as the fit curves but without the detector-induced broadening. For orthogonal polarization this results in a pure two-sided exponential decay in time.

We used the same set of parameters to model the central peak of the HOM interference data for parallel polarization shown in red in Fig.~\ref{fig:hybrid_hom}b). The model function now has an interfering component due to the coherent part of the quantum dot photons for which we use the coherence time $T_2=\SI{216(3)}{ps}$. As mentioned above, this is likely underestimating the actual coherence, but given the detector broadening, we cannot make any more accurate statements. In addition, this value is equivalent to the one used in Ref.~\cite{Polyakov11a} and thus helps the comparison. The interfering part results in a sharp dip at the center of the broader peak, for the ideal case with no detector time jitter. The latter washes this feature out resulting in the characteristic flattened peak. Using the definition of the coalescence probability \cite{Polyakov11a} $P_C:= (A_\perp-A_\|)/A_\perp$, where $A_\perp$ is the total number of coincidence counts in the central peak for orthogonal polarization and $A_\|$ the same for parallel polarization we find $P_C = 0.39(4)$. Using the $T_1$ and $T_2$ values from above we get a maximum possible value of $P_C^\mathrm{max} = 0.36(3)$, which is obviously underestimated because of the underestimated value of $T_2$.

We can improve on the coalescence by additional time filtering in post-processing at the cost of further reduced efficiency. For this purpose, we limit the acceptable detection events to three \SI{128}{ps}-wide time bins, i.e. \SI{384}{ps}, for each detector using the earliest photons to arrive after a laser pulse for both detectors. The respective time bins are in the light green shaded area in Fig.~\ref{fig:pulsehistograms}c) and d) for the two detectors $\mathrm D_1$ and $\mathrm D_2$, respectively. The filtering window, however, is limited in its effective sharpness by the finite detector time jitter. Still it will reject events that are far away from the zero time delay, which are for the quantum dot those that were emitted late and thus incoherently. Effectively this filtering thus increases the indistinguishability. This procedure results in the histogram of detection time differences shown in Fig.~\ref{fig:hybrid_hom}, which happens to only have counts in four discrete detection time differences. While we could expect there to be up to seven different bins with nonzero counts only four happen to contain any events. For this time-filtered data the coalescence probability is 0.63(5). Investigating other filtering windows we found that including only two more time bins brings us almost back to the unfiltered value of the coalescence probability. This post-selection has the same effect as a temporal shaping of the QD photons, as presented in Refs.~\cite{Huber15a,Ates13a}. In the case of our quantum dot-SPDC HOM interference, the temporal filtering discards photons that do not overlap in time, on top of removing incoherent photons from the QD luminescence.

\section{Discussion and Outlook}
To place these results in perspective we compare them with an earlier measurement using the same setup but with a \SI{30}{GHz} fiber-Bragg grating instead of the pulse stretcher for filtering, and to the results obtained in Ref.~\cite{Polyakov11a}. There the authors used an actively stabilized Fabry-P\'erot cavity for filtering the SPDC photons and a similar quantum dot, but with quasi-resonant higher shell excitation, which introduces more time jitter into the state preparation of the QD. In Table~\ref{tbl:comparison} we give a comparison of the most relevant parameters in the systems. The maximum theoretical coalescence value we calculated is the highest possible result given the different photon wavepacket shape or spectrum but without including the extra dephasing of the quantum dot photons, without postselection, i.e., this would be the result for a lifetime-limited QD for the given filters, without temporal post selection. The temporal post selection can go over this value, because only photons with a better temporal overlap are considered. In addition, the coherence times extracted from Michelson interferometer measurements underestimate the coalescence, so that even the unfiltered value can be slightly above.

\begin{table}
  \centering
  \begin{tabular}{lSSSSs}
    \toprule
    &\parbox{2cm}{Fiber Bragg grating} & \parbox{2cm}{Pulse stretcher} & \multicolumn{2}{c}{\parbox{2cm}{Ref.~\cite{Polyakov11a}}}      & {Unit} \\
    \midrule
    Max. theoretical coalescence & 17                    & 36                & \multicolumn{2}{c}{\tablenum{67}}                & \%     \\
    Measured raw coalescence     & 18(4)              & 39(4)          & \multicolumn{2}{c}{\tablenum{16(3)}}            & \%     \\
    Time-selected coalescence    & 35(9)              & 63(5)          & 61                                         & 75  & \%     \\
    Time window                  & 384                   & 384               & 290                                        & 140 & ps     \\
    Time selection efficiency    & 13.3                  & 12.6              & 25                                         & 10  & \%     \\
    $2T_1/T_2$                   & 3.0                  & 3.0              & \multicolumn{2}{c}{\tablenum{5.7}}              &        \\
    $\Delta\nu_\mathrm{SPDC}$    & 30                    & 7.7               & \multicolumn{2}{c}{\tablenum{0.9}}               & GHz    \\
    $\Delta\nu_\mathrm{QD}$      & 1.2                   & 1.2               & \multicolumn{2}{c}{\tablenum{1.1}}               & GHz    \\
    \bottomrule
  \end{tabular}

  \caption{Comparison of our results with an earlier measurement using only a fiber Bragg grating for filtering instead of the pulse stretcher, and with work by Polyakov et al.~\cite{Polyakov11a}. ``Maximum theoretical coalescence'' here is defined as the highest possible result given the different photon wavepacket shape or spectrum, but not including any extra dephasing, impurity, or other corrections to the Michelson interference coherence time.}
  \label{tbl:comparison}
\end{table}

Even with some time-filtering via a narrow coincidence window the coalescence stays far below unity. If we were to use our quantum dot and excitation method with bandwidth-matched Fabry-Perot filtering we would expect a coalescence fraction of 86~\%, even when taking the dephasing in the quantum dot into account. To get from this value to a high-fidelity quantum interface several things need to improve. Most importantly, one needs to use quantum dots that emit indistinguishable photons, i.e. with high purity, which has recently been achieved for InAs/GaAs quantum dots~\cite{Wang16a,Loredo16a}. While we assume that appropriate filtering of SPDC will still be possible, this remains costly in terms of efficiency to match the usual quantum dot spectral bandwidth ($\lesssim\SI{1}{GHz}$). An alternative approach is to pursue intracavity SPDC as discussed in the introduction, but we believe that a better option is to use even shorter lifetime quantum dots such as the GaAs/AlGaAs system, where quantum dots are grown by droplet epitaxy. Early results show radiative lifetimes of less than \SI{250}{ps}~\cite{Huber17a} and high indistinguishability, without any filtering or shaping.

We are confident that these early experiments like Ref.~\cite{Polyakov11a} or ours help identify systematic and practical problems with optical quantum interfaces. They show us which improvements are most likely to solve the problems towards a future quantum internet. Single spins in quantum dots may well serve as stationary qubits at least for local storage in quantum repeater nodes and thus with this work we get closer to solving the problem of how to entangle remote quantum dots using high quality entanglement from a spontaneous parametric down-conversion source.

\textit{Acknowledgments}\\
GSS acknowledges support from the PFC@JQI and TH acknowledges support through a JQI postdoctoral fellowship. AP acknowledges funding for part of this work by the Austrian Science Fund (FWF), project no.~V-375. This work was funded in part by the European Research Council (ERC), project ``\emph{EnSeNa}'' (257531). MP was supported in part by the Austrian Science Fund Doctoral Program ``\emph{Atoms, Light and Molecules}'' project no.~W-1259.

\bibliographystyle{iopart-num}
\bibliography{hybridhom}

\end{document}